\documentclass[conference]{IEEEtran}

\IEEEoverridecommandlockouts

\ifCLASSINFOpdf
  \usepackage[pdftex]{graphicx}
  \DeclareGraphicsExtensions{.pdf,.jpeg,.png}
  \graphicspath{{figs/}}
\else
\fi

\usepackage{cite}
\usepackage{amsmath}
\usepackage{siunitx} 
\usepackage{graphicx}
\usepackage{amssymb}
\usepackage{pifont}
\usepackage{multicol}
\usepackage{lipsum}
\usepackage{url}
\usepackage{mathtools}
\usepackage{enumitem}
\usepackage{xcolor}
\usepackage[linesnumbered,ruled,vlined]{algorithm2e}
\DontPrintSemicolon
\usepackage{algpseudocode}
\usepackage{multirow}

\usepackage[justification=centering]{caption}

\usepackage{hyperref}
\hypersetup{
  backref         = true ,
  pagebackref     = true ,
  hyperindex      = true ,
  colorlinks      = true ,
  breaklinks      = true ,
  urlcolor        = black,
  linkcolor       = red  ,
  bookmarks       = true ,
  bookmarksopen   = false,
  filecolor       = black,
  citecolor       = blue ,
  linkbordercolor = blue
}

\hypersetup{final}

\usepackage[font=footnotesize,skip=3pt]{caption}
\usepackage{xspace}

\newcommand{\HEPPO}{HEPPO-GAE\xspace}

\newcommand{\RL}{RL\xspace}

\begin{document}

\bstctlcite{IEEEetal:BSTcontrol}
\bstctlcite{IEEEdash:BSTcontrol}

\setlength{\abovedisplayskip}{3pt}

\renewcommand{\KwSty}[1]{\textnormal{\textcolor{blue!90!black}{\ttfamily\bfseries #1}}\unskip}
\renewcommand{\ArgSty}[1]{\textnormal{\ttfamily #1}\unskip}
\SetKwComment{Comment}{\color{green!50!black}// }{}
\renewcommand{\CommentSty}[1]{\textnormal{\ttfamily\color{green!50!black}#1}\unskip}
\newcommand{\assign}{\leftarrow}
\newcommand{\var}{\texttt}
\newcommand{\FuncCall}[2]{\texttt{\bfseries #1(#2)}}
\SetKwProg{Function}{function}{}{}
\renewcommand{\ProgSty}[1]{\texttt{\bfseries #1}}
\SetKwInOut{KwIn}{Input}
\SetKwInOut{KwOut}{Output}
\SetKwProg{Init}{Initialization}{}{}
\SetKwProg{DataIns}{Data Insertion}{}{}
\SetKwProg{CalcUpd}{GAE Calculation and In-Place Update}{}{}

\title{
\fontsize{19.71}{23.652}\selectfont \HEPPO: Hardware-Efficient Proximal Policy Optimization with Generalized Advantage Estimation
\vspace{-10pt}
}

\author{
	\IEEEauthorblockN{Hazem Taha and Ameer M. S. Abdelhadi}
	\IEEEauthorblockA{
	    \textit{McMaster University}\\
	    1280 Main St. W, Hamilton, Ontario L8S 4L8\\
	    \{tahah,ameer\}@mcmaster.ca
        \vspace{-20pt}
    }
}

\maketitle

\begin{abstract}
This paper introduces \HEPPO, an FPGA-based accelerator designed to optimize the Generalized Advantage Estimation (GAE) stage in Proximal Policy Optimization (PPO). Unlike previous approaches that focused on trajectory collection and actor-critic updates, \HEPPO addresses GAE's computational demands with a parallel, pipelined architecture implemented on a single System-on-Chip (SoC). This design allows for the adaptation of various hardware accelerators tailored for different PPO phases. A key innovation is our strategic standardization technique, which combines dynamic reward standardization and block standardization for values, followed by 8-bit uniform quantization. This method stabilizes learning, enhances performance, and manages memory bottlenecks, achieving a 4x reduction in memory usage and a 1.5x increase in cumulative rewards. We propose a solution on a single SoC device with programmable logic and embedded processors, delivering throughput orders of magnitude higher than traditional CPU-GPU systems. Our single-chip solution minimizes communication latency and throughput bottlenecks, significantly boosting PPO training efficiency. Experimental results show a 30\% increase in PPO speed and a substantial reduction in memory access time, underscoring \HEPPO's potential for broad applicability in hardware-efficient reinforcement learning algorithms.
\end{abstract}

\IEEEpeerreviewmaketitle

\vspace{-8pt}

\section{Introduction}

\textit{Reinforcement Learning (RL)} is a subset of machine learning where agents learn best behaviors by interacting with the environment. RL agents do not receive correct input/output pairs like in supervised learning; they find strategies through trial and error, guided by rewards. This approach has been crucial in addressing intricate decision-making challenges in various fields such as robotics \cite{levine2018rlrobotics} and strategic games like chess and Go \cite{silver2017rlgames}.

Proximal Policy Optimization (PPO) is a widely used reinforcement learning (RL) algorithm that optimizes policy directly through gradient ascent to maximize expected cumulative reward \cite{schulman2017proximal}. PPO enhances the stability of policy gradient methods by using a clipped objective function to ensure that policy updates are not excessively large, effectively addressing high gradient variance and instability. This approach eliminates the need for the computationally expensive second-order optimization step required by algorithms such as Trust Region Policy Optimization (TRPO), making PPO easier to implement and more computationally efficient \cite{schulman2015trust}. By preventing large updates, PPO maintains robustness and improves training stability of TRPO \cite{schulman2017proximal}. At the core of PPO, it effectively balances the need for exploration and exploitation by preventing large policy updates, ensuring stable and efficient learning. This makes PPO a robust and practical choice for a wide range of \RL tasks \cite{schulman2017proximal}.

\begin{algorithm}[h]
\begin{minipage}{\columnwidth}
\small
\KwIn{(1) initial policy parameters $\theta_0$,\\ (2) initial value function parameters $\phi_0$}
\KwOut{learned policy $\pi_{\theta}$}
\For{$k = 0, 1, 2, \ldots$}{
    \textbf{Collect} a set of trajectories $D_k = \{\tau_i\}$ by running policy
    
    ~~~~~~~~~ $\pi_k = \pi(\theta_k)$ in the environment.\;
    \textbf{Compute} rewards-to-go $\hat{R}_t$.\;
    \textbf{Compute} advantage estimates, $\hat{A}_t$, based on the current
    
    ~~~~~~~~~~~ value function $V_{\phi_k}$.\;
    \textbf{Update} the policy by maximizing the PPO-Clip objective
    \vspace{-6pt}
    \begin{multline*}
                    \theta_{k+1} = \arg\max_{\theta} \frac{1}{|D_k|T} \sum_{\tau \in D_k} \sum_{t=0}^{T}  \min ( \cr \frac{\pi_{\theta}(a_t|s_t)}  {\pi_{\theta_k}(a_t|s_t)} A^{\pi_{\theta_k}}(s_t, a_t), g(\epsilon, A^{\pi_{\theta_k}}(s_t, a_t)) ) ,
    \end{multline*}\;\vspace{-12pt}
    \footnotesize\Comment{typically via stochastic gradient ascent with Adam \cite{Kingma2015Adam}}
    \textbf{Fit} the value function by regression on mean-squared error
    \begin{equation*}
            \phi_{k+1} = \arg\min_{\phi} \frac{1}{|D_k|T} \sum_{\tau \in D_k} \sum_{t=0}^{T} \left( V_{\phi}(s_t) - \hat{R}_t \right) ^2,
    \end{equation*}\;\vspace{-14pt}
    \footnotesize\Comment{typically via some gradient descent algorithm}
}
\caption{PPO Algorithm}
\label{alg:PPOalgorithm}
\end{minipage}
\vspace{-4pt}
\end{algorithm}

In \autoref{alg:PPOalgorithm}, $\theta$ represents policy parameters, and $\phi$ represents value function parameters. The policy $\pi_{\theta}$ maps states to actions, while $\pi_{\theta_k}$ denotes the policy at the $k$-th iteration. Rewards-to-go $\hat{R}_t$ are the sum of future rewards from time step $t$, and advantage estimates $\hat{A}_t$ measure how much better an action is compared to the average action at a given state. The function $g(\epsilon, \hat{A}_t)$ clips the probability ratio to prevent large updates.

An essential component of PPO is the computation of advantage estimates. The advantage function, $A^\pi(s, a)$, measures how much better taking action $a$ in state $s$ compared to the average action taken in state $s$ under policy $\pi$, namely,
\begin{equation}
A^\pi(s, a) = Q^\pi(s, a) - V^\pi(s),
\end{equation}
where \( Q^\pi(s, a) \) is the state-action value function, and \( V^\pi(s) \) is the state value function.

To compute these advantage estimates effectively, Generalized Advantage Estimation (GAE) is used. GAE helps reduce the variance of the advantage estimates, leading to more stable and efficient policy updates \cite{schulman2016high}.

\subsection{Background on Generalized Advantage Estimation (GAE)}
GAE addresses the variance-bias tradeoff in policy gradient methods for RL by using value functions to estimate the advantage function more accurately, at the cost of introducing some bias. The key idea is to use an exponentially-weighted estimator of the advantage function, analogous to the TD(\(\lambda\)) method \cite{schulman2016high}.
An estimation can be derived by utilizing the temporal-difference (TD) residual, \( \delta V_t \),
\begin{equation}
\delta V_t = r_t + \gamma V(s_{t+1}) - V(s_t),
\end{equation}
whereas the GAE is an exponentially-weighted average of $k$-step advantage estimators:
\begin{equation}
\hat{A}^{GAE(\gamma, \lambda)}_t = \sum_{l=0}^{\infty} (\gamma \lambda)^l \delta V_{t+l}.
\end{equation}
Alternatively, this can be computed sequentially, namely,
\begin{equation}
\label{equation_adv}
A_t^{GAE} = \delta_t + (\lambda \gamma) A_{t+1}^{GAE}.
\end{equation}

GAE allows for direct computation of advantages, handling reward delays and noisy rewards more effectively, where Rewards-to-Go are defined as
\begin{equation}
\text{Rewards-to-Go} = V_t + \hat{A}^{GAE}_t.
\end{equation}

GAE is used in PPO for policy updates as it provides low-variance and high-bias advantage estimates. This involves
\begin{enumerate}
    \item Collecting trajectories using the current policy.
    \item Computing the advantages and rewards-to-go for each state-action pair.
    \item Using the computed advantages to update the policy using the PPO objective.
\end{enumerate}

This combination allows PPO to leverage GAE's strengths, resulting in improved performance on complex RL tasks.

\subsection{Challenges and Limitations in PPO Acceleration}

\begin{figure}[hb]
\centering
\includegraphics[width=0.5\textwidth]{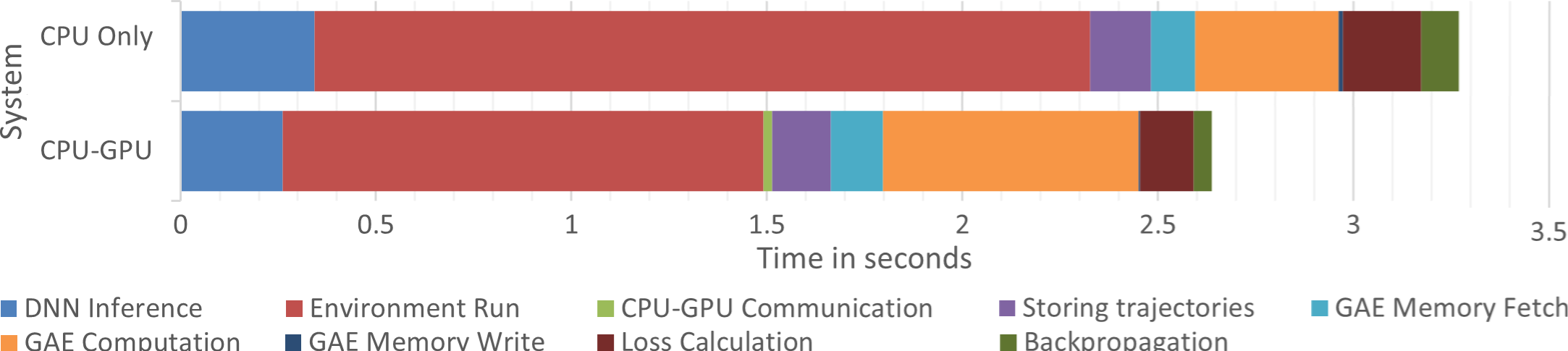}
\caption{Time Profiling of PPO Iteration over Different Systems}
\label{fig:time_profiling}
\end{figure}

\autoref{tab:time_profiling} and \autoref{fig:time_profiling} highlight the time taken by different components of the PPO algorithm in both CPU-only and CPU-GPU systems. The profiling was conducted on a high-performance system with 32 Intel(R) Xeon(R) Silver 4216 CPU cores @ 2.10GHz and a Tesla V100-SXM2-32GB GPU, using the Humanoid environment by Gymnasium. 

The data reveals that the environment run and the GAE computation phase consume a significant portion of the processing time (47\% and 30\% respectively in CPU-GPU systems). Knowing that environments are typically high-level code compiled to run on commodity CPUs and are independent of the RL algorithm, it’s not feasible to build custom hardware that can accelerate different environments. 

In our work, we focused on exploiting the PPO characteristics to accelerate the algorithm itself. Developing custom hardware for the GAE phase and potentially other heavy components of PPO and executing these computations on an SoC would reduce the communication overhead between different systems like CPU, GPU, and DRAM. Replacing DRAM with on-chip memory for these operations would further decrease latency and improve data throughput, leading to significant performance gains in PPO training.

\begin{table}[ht]
\caption{Time Profiling of PPO Iteration over Different Systems\\(as percentages of total time)}
\centering
\begin{tabular}{|p{35pt}||l|c|c|}
\hline
\textbf{Phase} & \textbf{Sub-Phase} & \textbf{CPU-GPU} & \textbf{CPU Only} \\ \hline\hline
\multirow{4}{*}{\textbf{\shortstack{Trajectory\\Collection}}} & DNN Inference & 9.92\% & 10.46\% \\ \cline{2-4}
& Environment Run & 46.58\% & 60.71\% \\ \cline{2-4}
& CPU-GPU Communication & 0.85\% & NA \\ \cline{2-4}
& Storing Trajectories & 5.73\% & 4.75\% \\ \hline
\multirow{3}{*}{\textbf{GAE}} & GAE Memory Fetch & 5.00\% & 3.49\% \\ \cline{2-4}
& GAE Computation & 24.79\% & 11.23\% \\ \cline{2-4}
& GAE Memory Write & 0.17\% & 0.32\% \\ \hline
\multirow{2}{*}{\textbf{\shortstack{Network\\Update}}} & Loss Calculation & 5.21\% & 6.10\% \\ \cline{2-4}
& Backpropagation & 1.77\% & 2.95\% \\ \hline
\end{tabular}
\label{tab:time_profiling}
\end{table}

\subsection{Related Work}

This section discusses key contributions in hardware acceleration for RL, aligning with our work on \HEPPO by addressing the computational demands of RL algorithms through innovative hardware designs, quantization techniques, and memory management strategies.

Krishnan et al. introduced an RL training paradigm using 8-bit quantized actors to accelerate data collection without compromising learning convergence, achieving a 1.5× to 5.41× speedup and reducing carbon emissions by 1.9× to 3.76× compared to full-precision training \cite{krishnan2022quarl}. Yang et al. employed fixed-point data types and arithmetic units for both training and inference, demonstrating a training throughput of 25293.3 inferences per second (IPS), 2.7 times higher than a CPU-GPU platform, and an energy efficiency of 2638.0 IPS/W, 15.4 times more energy-efficient than a GPU \cite{yang2022fixar}. These studies highlight the effectiveness of quantization techniques in enhancing RL training speed and energy efficiency.

Meng et al. (2020) targeted the inference and training phases of the PPO algorithm on a CPU-FPGA heterogeneous platform, achieving throughput improvements of 2.1×–30.5× over CPU-only implementations and 2×–27.5× over CPU-GPU implementations \cite{ppo_paper}. Specific benchmarks showed a 23.5\% increase in throughput for Hopper and a 21.2\% increase for Humanoid with data layout optimization. Load balancing optimization led to improvements ranging from 9.3\% to 28.3\% in overall running average throughput.

Weng et al.'s EnvPool addresses the bottleneck of slow environment execution in RL training systems using a C++ thread pool-based executor engine, achieving 1 million frames per second for Atari environments and 3 million frames per second for MuJoCo environments on a NVIDIA DGX-A100 with 256 CPU cores \cite{weng2022envpool}. Dalton et al.'s CuLE platform leverages GPU parallelization to run thousands of Atari game environments simultaneously, achieving up to 155 million frames per hour \cite{dalton2020cule}. Liang et al. introduced a GPU-accelerated RL simulator using NVIDIA Flex, achieving substantial improvements in training complex RL tasks and offering significant scaling benefits with multi-GPUs \cite{liang2018gpu}. These works underscore the critical role of efficient environment simulation in enhancing RL training performance.

To the best of our knowledge, we are the first to specifically target the optimization of the critical GAE step in PPO, which constitutes around 30\% of the total processing time in CPU-GPU systems. This work addresses this gap by focusing on the computational demands of GAE, offering a significant contribution to the field of RL hardware acceleration.

\subsection{Paper Contribution}

Major contributions and innovations of this paper are
\begin{itemize}

    \item Enabling the integration of multiple custom hardware components, memory, and CPU cores on a single system-on-chip (SoC) architecture, accommodating all phases of PPO from environment simulation to GAE computation. This reduces communication overhead and enhances data throughput and system performance.

    \item Introducing dynamic standardization for rewards and block standardization for values. This technique stabilizes learning, enhances training performance, and manages memory efficiently, reducing memory usage by 4x and increasing cumulative rewards by 1.5x.

    \item A parallel processing system that processes trajectories concurrently, employing a $k$-step lookahead approach for optimized advantage and rewards-to-go calculations. Our pipelined Processing Element (PE) can handle 300M elements per second, decimating the delay of GAE calculation and reducing PPO time by approximately 30\%.

    \item A memory layout system that organizes rewards, values, advantages, and rewards-to-go on-chip for faster access. Using dual-ported Block RAM (BRAM) to implement a FILO storage mechanism, this system provides the required throughput each cycle, allowing overwriting of the same memory locations for efficient data handling. 

    \item In-depth time profiling for the PPO algorithm revealing that GAE computation is a major contributor to processing time, accounting for 30\% in CPU-GPU systems.

\end{itemize}

\section{Algorithm Modification}
\label{sec:Algorithm_Modification}

We aim to achieve an optimally-reduced version of the PPO algorithm that can closely resemble the training behavior of the original algorithm, while allowing rescaling the input data to the GAE calculation phase. This guarantees that any computation done to the input data will be independent of the used environments and hyperparameters as all inputs are re-distributed evenly. To achieve our goal, several modifications have been proposed and investigated as follows.

\subsection{Dynamic Standardization of Rewards}

The motivation behind standardizing the rewards (and later values) is to have a consistent and predictable distribution in which we can perform quantization. Applying traditional standardization techniques has experimentally shown to cause training divergence. This is mainly because these methods independently alter the distribution of rewards within each training epoch, disrupting the relative differences in reward distributions between epochs and equalizing short-term and long-term rewards, misleading the training.

To solve this problem, a novel standardization technique has been developed and coined the name Dynamic Standardization. The idea is that at each training epoch, reward standardization shall be conducted while accounting for all previously attained rewards. As it will be computationally and memory intensive to store and reprocess all the rewards across training, a more efficient way is to store a running mean and running standard deviation that gets updated every epoch with the new reward.

To update the running mean with every new reward, we follow the equation
\begin{equation}
\resizebox{.91\columnwidth}{!}{
    $\text{RunningMean}_n = \text{RunningMean}_{n-1} + \frac{r_n - \text{RunningMean}_{n-1}}{n}$,
}
\end{equation}

where $n$ is the total number of rewards processed so far, $r_n$ is the $n$-th reward, and $\text{RunningMean}_n$ is the running mean calculated up to the $n$-th reward.

As for the running standard deviation, inspired by Welford's algorithm \cite{welford1962} \cite{knuth1998} for dynamically calculating variance over multiple iterations, the running variance for each new data point has been computed as follows.
\begin{enumerate}
    \item Initialize \( M_0 \) and \( S_0 \) to 0.
    \item For each new reward \( r_n \)
    \begin{equation}
        M_n = M_{n-1} + \frac{(r_n - M_{n-1})}{n}, \quad\quad\quad\text{and}
    \end{equation}
    \begin{equation}
        S_n = S_{n-1} + (r_n - M_{n-1}) \times (r_n - M_n).
    \end{equation}
    \item The running standard deviation after \( n \) rewards is then
    \begin{equation}
        \text{RunningSTD}_n = \sqrt{\frac{S_n}{n}},
    \end{equation}
\end{enumerate}

where $M_n$ is the running mean after $n$ rewards and $S_n$ is the cumulative value used for calculating variance.

\subsection{Block Standardization of Values}
\begin{figure}[b]
\centering
\includegraphics[width=0.5\textwidth]{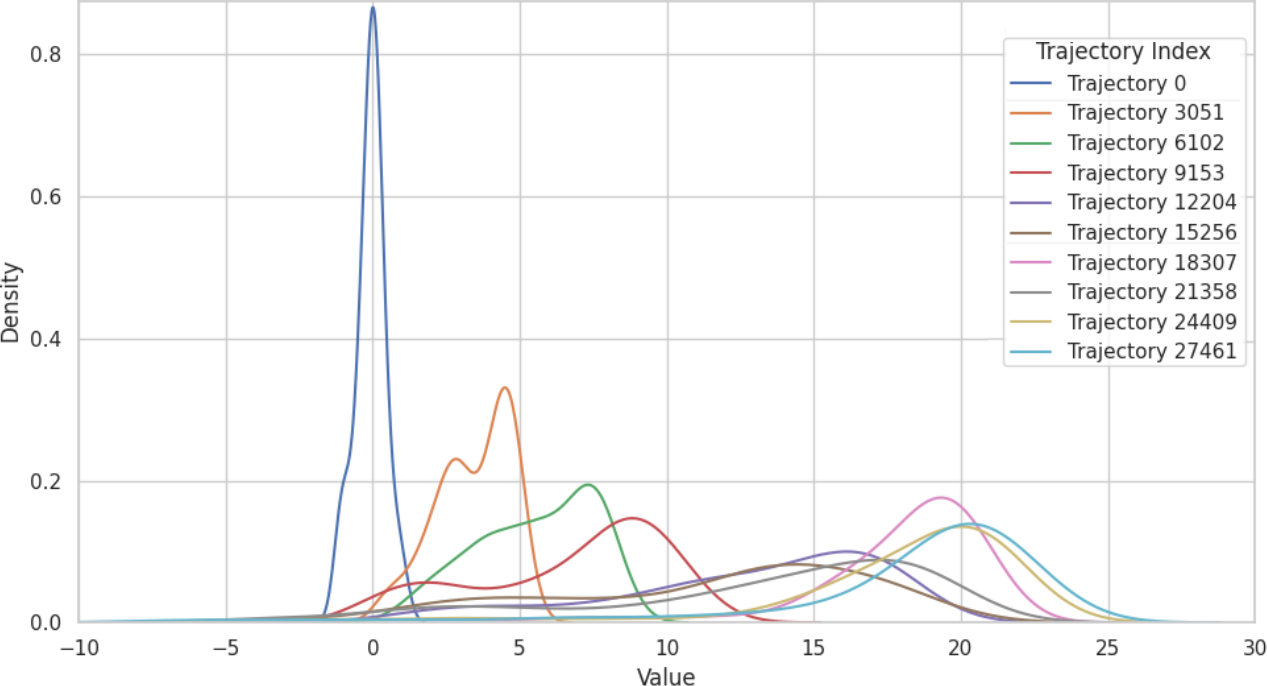}
\caption{Distribution of Value Across Collected Trajectories}
\label{fig:values_distribution}
\end{figure}

Unlike rewards, the values are outputs of a trainable Neural Network (critic) that evolves differently over time and exhibits varying distributions. This observation is illustrated in \autoref{fig:values_distribution}, which shows the distribution of values across a selected set of trajectories during training. 

Dynamic standardization of values was unsuccessful as it affected the loss calculations. Instead, a more adaptable standardization method is required to handle these variations effectively while keeping a history of their original distribution to project them back in place. To address this, we propose a block standardization technique that quantizes values in batches. The steps involved in this process are as follows:

\begin{enumerate}
    \item \textbf{Batch Collection:} Collect a batch of values from multiple trajectories.
    \item \textbf{Compute Statistics:} Calculate the mean ($\mu_v$) and standard deviation ($\sigma_v$) for each batch.
    \item \textbf{Standardization:} Scale values to have a mean of zero and a standard deviation of one by subtracting each the mean $\mu_v$ from each element in the block and then dividing by standard deviation $\sigma_v$.
    \item \textbf{Uniform Quantization:} Quantize standardized values uniformly, storing them with $\mu_v$ and $\sigma_v$.
    \item \textbf{Reconstruction:} De-quantize and convert values back to the original scale using the stored statistics.
\end{enumerate}

This method leverages the similar distribution of trajectories collected at the same point in training, allowing for adaptive quantization based on the actual mean and standard deviation at that moment. The effectiveness of this method was validated through a series of experiments, demonstrating its robustness to shifts in training dynamics and its efficiency in utilizing memory bandwidth.

\subsection{Quantization of Rewards and Values}

Due to the memory bottlenecks discussed in \autoref{sec:DataLayout}, it is impractical to use 32-bit floating-point representation for each element in the Rewards and Values vectors. Instead, we adopt a quantization strategy tailored separately for rewards and values.

\subsubsection{Quantization of Rewards}
After applying dynamic standardization, rewards are centered around zero with a unit standard deviation. They are then uniformly quantized using $n$-bit codeword, mapping continuous values to discrete levels. During reconstruction, rewards are retrieved from memory, de-quantized, and used in their standardized form. Experimental testing showed that leaving the rewards in their standardized form enhances the cumulative rewards by around 50\% as shown in section \autoref{sec:Results}.

\subsubsection{Quantization of Values}
Similar to the Quantization of Rewards, After applying block standardization, the values are uniformly quantized using $n$-bit codeword. During reconstruction, we also fetch and de-quantize the values. However, the main difference is that we have to do a final de-standardization step shifting the distribution back to its original form. This is done by multiplying the elements in $v$ back by the stored standard deviation $\sigma_v$ and then adding the mean $\mu_v$.

\section{\HEPPO Architecture Details}

\begin{figure}[b]
\centering
\includegraphics[width=\columnwidth]{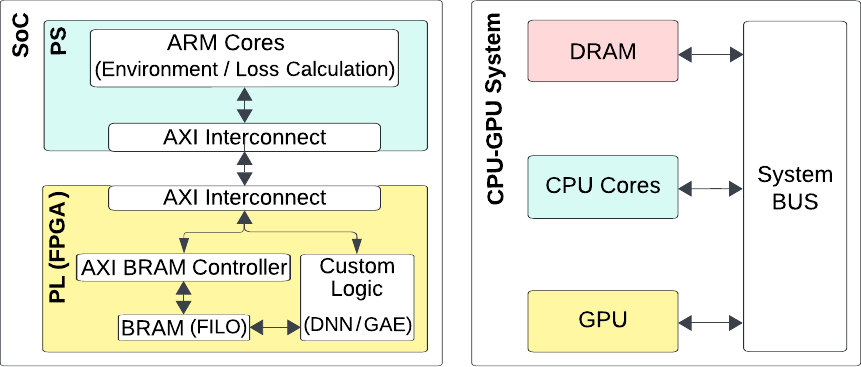}
\caption{Comparison of SoC Architecture and Traditional CPU-GPU System}
\label{fig:heppo_soc}
\end{figure}

The proposed architecture integrates the whole PPO pipeline in a single SoC, reducing latency and communication overhead compared to traditional CPU-GPU systems. \autoref{fig:heppo_soc} illustrates the connections between the Processing System (PS), Programmable Logic (PL), and BRAM within the SoC.

\subsection{Data Flow, Processing, and Efficiency}

The PS access the BRAM for reading and writing via the AXI Interconnect, ensuring seamless data exchange with the custom logic in the PL. This integration keeps critical data on-chip, reducing the need to access external DRAM and enhancing throughput. Unlike traditional CPU-GPU systems which require frequent data transfers between the CPU, GPU, and DRAM, leading to higher latency and communication overhead.
Detailed Processing Stages are as follows.
\begin{itemize}
\item \textit{Data Preparation and Initiation:} Processed data is stored in BRAMs, and the PS communicates to the FPGA with an initiate signal.
\item \textit{Advantages and RTGs Calculation:} The FPGA fetches the data, performs de-quantization, calculates advantages and rewards-to-go (RTGs), writes back, and sends a completion signal back to the PS.
\item \textit{Actor-Critic Losses Calculation:} The PS retrieves the computed advantages and RTGs from the BRAMs and calculates actor-critic losses.
\item \textit{Back Propagation and Networks Update:} The PS sends the losses to the FPGA to perform backpropagation to update the neural networks. Updated network parameters are then used for subsequent iterations.
\end{itemize}

\subsection{Advantage Estimate Decomposition and k-step Lookahead}
In \RL, the advantage estimate \( A(t) \) is a critical component for policy improvement. Using \autoref{equation_adv}, we can decompose the advantage estimate calculation as shown in \autoref{tab:advDecomp}.

\begin{table}[t]
    \caption{Decomposition of advantage estimates for different $t$ values}
    \centering
    \begin{tabular}{|l|lll|}
        \hline
        $\mathbf{t}$     & \multicolumn{3}{c|}{$\mathbf{\hat{A}_{t}}$}\\
        \hline\hline
        $T$     & $\hat{A}_T$       & $=$ & $\delta_T$\\
        \hline
        $T-1$   & $\hat{A}_{T - 1}$ & $=$ &  $C\delta_{T} + \delta_{T-1}$\\
                &                   & $=$ &  $C\hat{A}_{T} + \delta_{T-1}$ \\
        \hline
        $T-2$   & $\hat{A}_{T-2}$   & $=$ &  $C^2\delta_{T}  + C\delta_{T-1} + \delta_{T-2}$\\
                &                   & $=$ &  $C^2\hat{A}_{T} + C\delta_{T-1} + \delta_{T-2}$\\
        \hline
        $T-3$   & $\hat{A}_{T-3}$   & $=$ &  $C^3\delta_{T} + C^2\delta_{T-1} + C\delta_{T-2} + \delta_{T-3}$\\
                &                   & $=$ &  $C^2\hat{A}_{T-1} + C\delta_{T-2} + \delta_{T-3}$\\
                &                   & $=$ &  $C^3\hat{A}_{T} + C^2\delta_{T-1} + C\delta_{T-2} + \delta_{T-3}$\\
        \hline
    \end{tabular}
    \label{tab:advDecomp}
\end{table}

\noindent where $C$ is a constant defined as $\gamma \cdot \lambda$.

This decomposition shows how each advantage estimate depends on future values. An efficient way that developers use is to compute the estimates in revere order from $t = N$ to $t = 1$ to avoid recalculating the same terms multiple times, thus saving computational resources. 

\begin{figure}[b]
\centering
\includegraphics[width=\columnwidth]{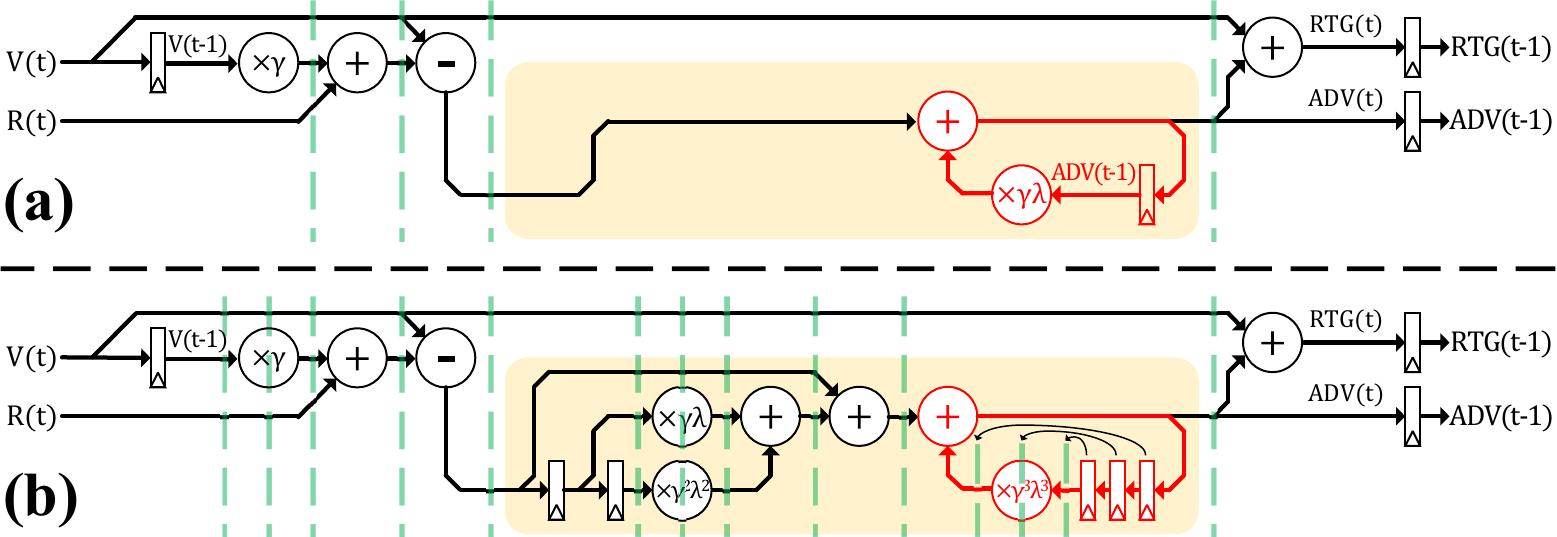}
\caption{Pipelining of the GAE unit: (a) possible pipelines (in dashed-green) are limited due to the logic loop (in red); (b) A 3-step Look-ahead (highlighted in orange) is applied to enable pipelining within the logic loop.}
\label{fig:Piplined}
\end{figure}

\textbf{\textit{Single Cycle Implementation and Pipelining}}:
The single-cycle GAE unit can be represented with potential pipelines highlighted in dashed green, as shown in \autoref{fig:Piplined}.
In this implementation, various stages of the computation are pipelined to improve efficiency. However, when we attempt to pipeline the feedback loop (highlighted in red), it introduces bubbles into the system. These bubbles represent idle states where the pipeline must wait for data from previous stages, severely reducing efficiency.

\textbf{\textit{k-step Lookahead Solution}}:
The k-step lookahead method addresses this inefficiency by introducing registers (delays) in the feedback loop. This approach can be explained through the modified advantage estimate calculations for 2-step lookahead
    \begin{equation}
        \hat{A}_{t} = C^2 \hat{A}_{t+2} + C \delta_{t+1} + \delta_{t}    ,
    \end{equation}
and for 3-step lookahead,
    \begin{equation}
        \hat{A}_{t} = C^3 \hat{A}_{t+3} + C^2 \delta_{t+2} + C \delta_{t+1} + \delta_{t}.
    \end{equation}

\autoref{fig:Piplined} illustrates the implementation of a 3-step lookahead. As shown, three registers are added to the feedback loop to apply 3-step lookahead transformation (highlighted in yellow). The added registers on the feedback loop can be moved inside the multiplier, enabling embedding pipelined multiplier through DSP blocks. This solution enables a fully pipelined processing, eliminating compute bubbles in the feedback loop. The following is a general equation for the k-step lookahead,
\[
\hat{A}_t = C^k \hat{A}_{t+k} + \sum_{i=0}^{k-1} C^{(k-1)-i} \delta_{t+i},
\]
facilitating the incorporation of additional registers on the feedback loop, and a more profound pipelining of the multiplier. Although the 2-step lookahead transformation is satisfactory for enabling our system to operate at the highest frequency and attain the peak performance, alternative systems, notably those with wider data formats, might necessitate more pipeline stages to operate at the maximum achievable frequency.

\subsection{System Architectue}

\label{sec:Architectural Details of HEPPO}

\begin{figure}[b]
\centering
\includegraphics[width=.97\columnwidth]{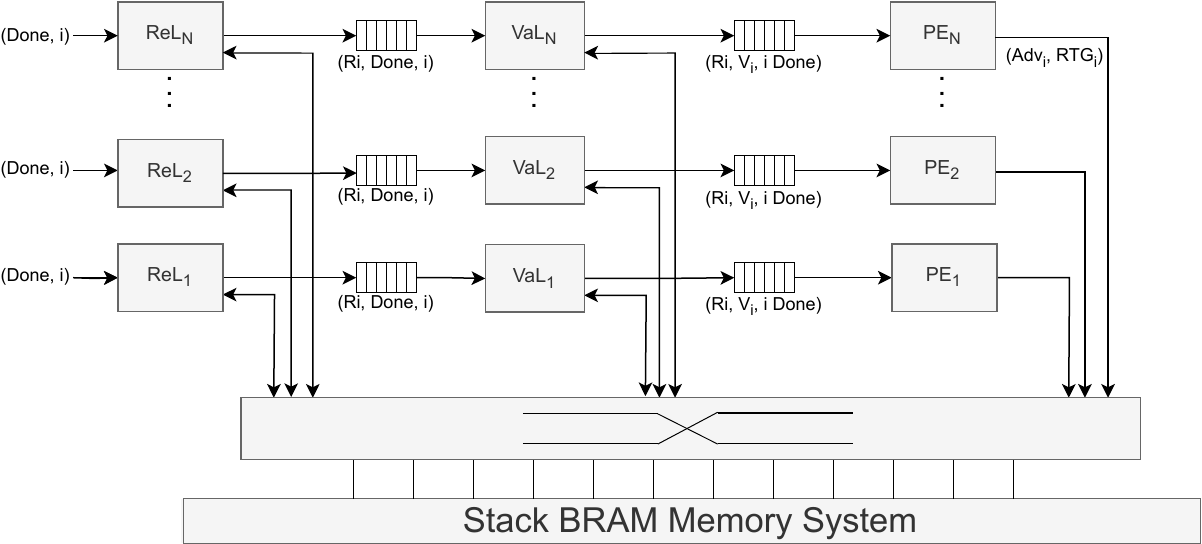}
\caption{\HEPPO architecture consisting of Rewards Loaders (ReLs), Values Loaders (VaLs), PEs, system crossbar, and BRAM stack memory}
\label{fig:Parallel Units}
\end{figure}

\autoref{fig:Parallel Units} (a) shows the micro-architecture of \HEPPO, which consists of Rewards Loaders (ReLs), Values Loaders (VaLs), and compute Processing Elements (PEs) forming a one-dimensional systolic array with N rows. 

\textbf{Parallelization:} Rows in the systolic array run concurrently and independently, each processing distinct vectors from different agents assigned by a round-robin fashion. When one row finishes, it gets a new set of vectors. This parallel architecture enhances \HEPPO's efficiency and scalability. While the BRAM stack memory enables substantial data transfers, a crossbar network ensures robust connections between ReLs, VaLs, and PEs to the BRAM stack memory.

\textbf{Data flow:} Each ReL reads element \( R_{i} \) from the rewards vector and sends it with index \( i \) and the signal Done to VaL. VaL fetches the corresponding \( i \)-th value \( V_{i} \) and sends \( R_{i} \), \( V_{i} \), \( i \), and Done to the PEs. The PE calculates the Advantage Estimate (Adv) and Rewards-to-Go (RTG) and writes them back to the main memory at index \( i \).

\section{Data Layout}
\label{sec:DataLayout}

To enhance the efficiency of the Proximal Policy Optimization (PPO) algorithm, we propose a memory layout that organizes rewards, values, advantages, and rewards-to-go on-chip for faster access. This layout groups data from different trajectories with the same timestep into memory blocks, enabling simultaneous retrieval and processing. Additionally, it employs a First-In-Last-Out (FILO) storage mechanism to align with the backward iteration required for GAE calculations. This section details the memory organization, access patterns, and bandwidth considerations.

\subsection{Memory Bandwidth Bottleneck}

In a typical large RL setup with 64 trajectories and 1024 timesteps, both rewards and values are stored in 32-bit floating-point format. The required memory per timestep for 64 trajectories (128 elements) is 512 bytes.

For parallel processing, these 512 bytes need to be fetched from memory per clock cycle. Assuming a typical DDR4 3200 bandwidth of 25~GB/s and a clock frequency of 300~MHz, the available bandwidth per cycle is calculated as $\text{Bandwidth per cycle} = \frac{25 \times 10^9\, \text{bytes/s}}{300 \times 10^6\, \text{cycles/s}} = 83.3\, \text{bytes/cycle}$.

This results in a shortfall of 428.7 bytes per cycle. Clearly, DRAM cannot supply enough data to sustain 64 parallel processing elements (PEs), severely limiting parallelization. To overcome this bottleneck, we store data in on-chip dual-port Block RAM (BRAM), which meets the required 512 bytes per cycle, ensuring high-throughput parallel processing.

\subsubsection{Memory Block Layout}

The memory layout is structured as 2D arrays, with dimensions representing timesteps and trajectories. Each memory block stores specific data (rewards, values, advantages, or rewards-to-go) indexed by timestep and trajectory. This layout enables parallel processing of different trajectories using the same fetched block which improves efficiency and lower memory accesses.

\begin{figure}[b]
    \centering
    \includegraphics[width=\columnwidth]{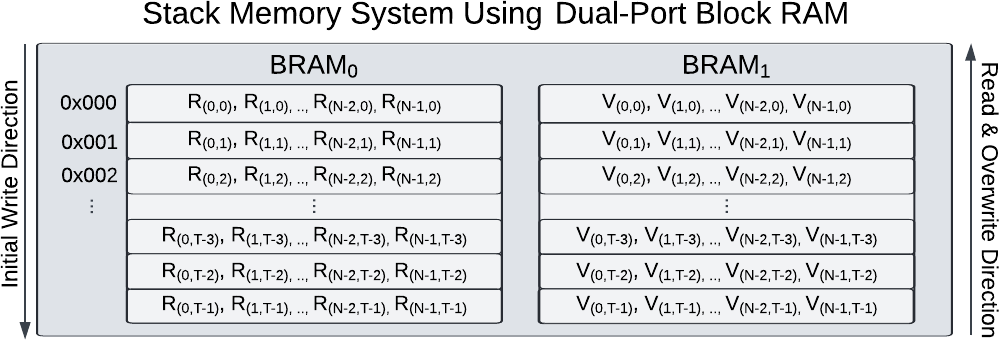}
    \caption{Dual-Port Block RAM Stack Memory System}
    \label{fig:memory_layout}
\end{figure}

\autoref{fig:memory_layout} illustrates the dual-port Block RAM (BRAM) stack memory system, consisting of:
\textbf{BRAM\textsubscript{0}}, which stores rewards $R_{i,j}$, and \textbf{BRAM\textsubscript{1}}, which stores values $V_{i,j}$, both from different trajectories \(i\) at the same timestep \(j\).

\subsubsection{FILO Storage Mechanism}

The FILO storage mechanism uses a stack-based structure to store rewards and values:
\begin{itemize}
    \item \textit{Push Operation}: Rewards and values are pushed onto the stack at each timestep.
    \item \textit{Pop Operation}: Rewards and values are popped from the stack during GAE calculation, starting from the last timestep and iterating backward.
\end{itemize}

\subsubsection{In-Place Updates and Dual-Port Memory}

The system uses dual-port memory for simultaneous read and write operations, enabling in-place updates where advantages and rewards-to-go can overwrite the original rewards and values reducing memory usage by half.

\autoref{alg:gae_stack_memory} is implemented to manage the FILO stack structure in BRAM, ensuring efficient data access patterns compatible with the hardware architecture.

This proposed design ensures fast data retrieval and processing that allows continuous feeding of data to the PEs, keeping them always busy.

\begin{algorithm}[t]
\caption{\fontsize{9.65}{10}\selectfont PPO/GAE Memory Layout and Processing}
\label{alg:gae_stack_memory}
\Init{}{
    \fontsize{9.55}{10}\selectfont\textbf{Initialize} memory blocks RMB, VMB, AMB, RTGMB
}
\DataIns{}{
    \ForEach{timestep $t$}{
        \ForEach{trajectory $i$}{
            \textbf{Push} $reward[i][t]$ into $\text{RMB}[t][i]$\;
            \textbf{Push} $value[i][t]$ into $\text{VMB}[t][i]$\;
        }
    }
}
\CalcUpd{}{
    \ForEach{timestep $t$, backward}{
        \ForEach{trajectory $i$}{
            \textbf{Retrieve} $reward$ from $\text{RMB}[t][i]$\;
            \textbf{Retrieve} $value$ from $\text{VMB}[t][i]$\;
            \textbf{Compute} advantage and reward-to-go\;
            \textbf{Store} advantage in $\text{AMB}[t+1][i]$\;
            \textbf{Store} reward-to-go in $\text{RTGMB}[t+1][i]$\;
        }
    }
}
\end{algorithm}

\section{Experimental Results}
\label{sec:Results}

The results of our work are divided across multiple axes. In this section, we will discuss each axis and how they can affect the overall acceleration of the PPO algorithm.

\subsection{Dynamic Rewards Standardization}

It's important to note that, in most PPO implementations, the final calculated advantage vector is standardized to stabilize gradient updates and ensure smoother and more consistent training. This practice has become widely adopted due to its positive impact on training dynamics, as highlighted in various implementations and community discussions \cite{PPOImplementation2021, OpenAIGitHub2018}.

\begin{figure}[b]
\centering
\includegraphics[width=\columnwidth]{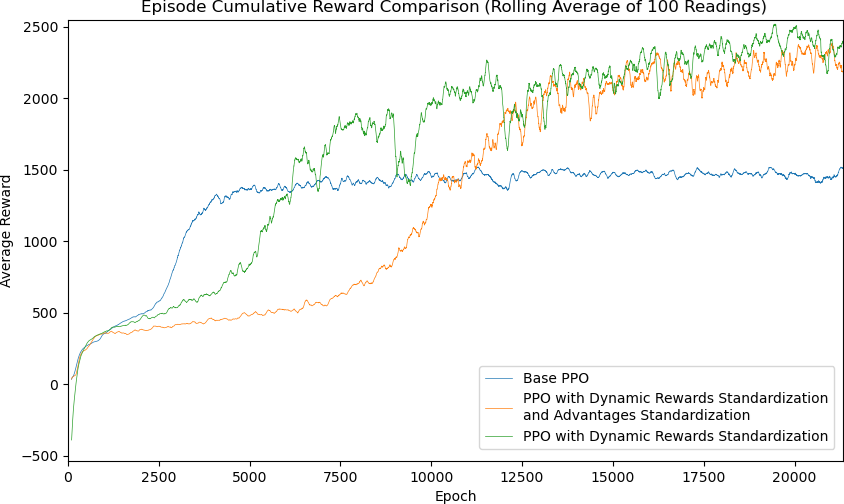}
\caption{Comparative Analysis of Cumulative Rewards Between Original PPO and Modified PPO with Dynamic Standardization}
\label{fig:Dynamic_Standerdization}
\end{figure}

\autoref{fig:Dynamic_Standerdization} shows training outcomes in the Humanoid environment (Gymnasium toolkit). Our PPO version (with and without standardized advantages) achieved over 1.5x increase in cumulative rewards compared to the original PPO, continuing to improve after the original PPO plateaued. This improvement was consistent across MuJoCo and Atari environments, confirming our modification benefits both hardware and training. 

\subsection{Quantization of Rewards and Values}

\begin{figure}[b]
    \centering
    \includegraphics[width=\columnwidth]{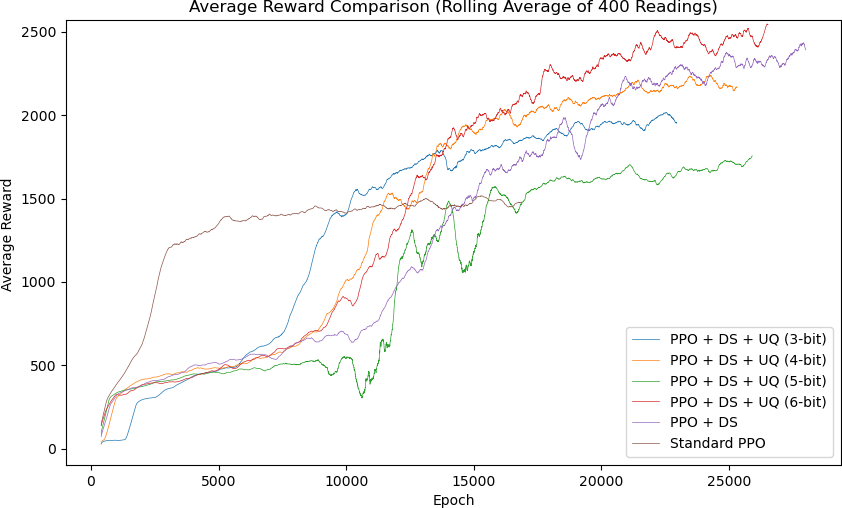}
    \caption{Uniform Quantization (3-6 bits) of Rewards}
    \label{fig:UniformQuantization1}
\end{figure}

\begin{figure}[b]
    \centering
    \includegraphics[width=\columnwidth]{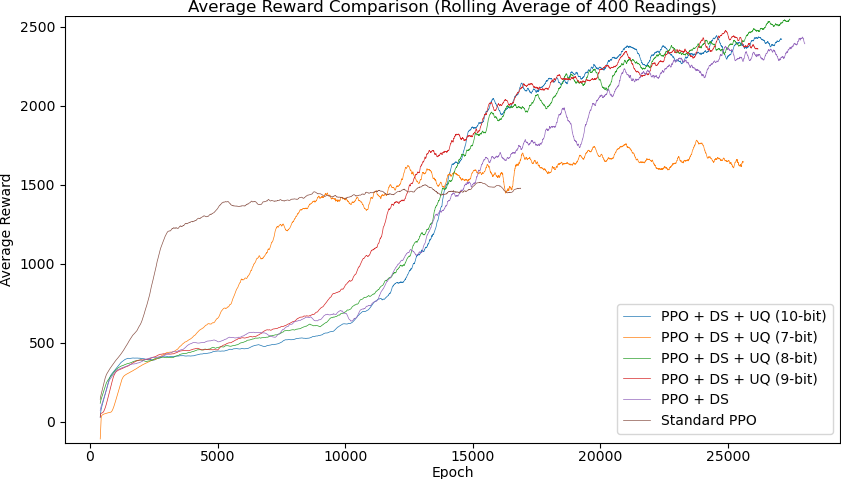}
    \caption{Uniform Quantization (7-10 bits) of Rewards}
    \label{fig:UniformQuantization2}
\end{figure}

\textbf{Optimal Quantization Size:} Detailed investigation, shown in \autoref{fig:UniformQuantization1} and \autoref{fig:UniformQuantization2} shows that quantizing using 5 and 7 bits performed the poorest followed by 3, 4 which are near to the baseline (PPO + DS). Finally, quantizing with 6, 8 to 10 performed equally higher than the baseline. The reason why using 5 and 7 bits performed worse than 3 and 4 in some of the trials and better in others is most likely due to the inherent variance of the policy gradient algorithm and the probabilistic nature of RL. To avoid this unstable region, it was concluded through all trials that 8 bits and above can be seen as a threshold for a stable uniform quantization that archives high performance.

\subsection{Summary of Rewards and Values Quantization Approaches}

To summarize the effects of various quantization approaches on PPO performance, we conducted several experiments with their configurations shown in \autoref{tab:experiments}. 

\begin{itemize}
    \item \textbf{Experiment 1}: Baseline PPO without quantization.
    \item \textbf{Experiment 2}: Dynamic standardization of rewards.
    \item \textbf{Experiment 3}: Both rewards and values are standardized and uniformly and quantized by block to 8-bit codewords.
    \item \textbf{Experiment 4}: Both rewards and values are standardized and uniformly and quantized by block, with rewards kept in standardized form throughout computations. 
    \item \textbf{Experiment 5}: Dynamic standardization applied to rewards and block approach to values.
\end{itemize}

These experiments highlight the importance of dynamic standardization and appropriate quantization techniques in improving PPO training efficiency. Experiment 4 performance was poor, indicating that simply keeping rewards standardized does not enhance performance but keeping them in the dynamically standardized form does. Experiment 5, combining dynamic standardization for rewards and Block Quantization for values, showed the best performance, emphasizing the significance of the way rewards are standardized.

\begin{figure}[b!]
    \centering
    \includegraphics[width=\columnwidth]{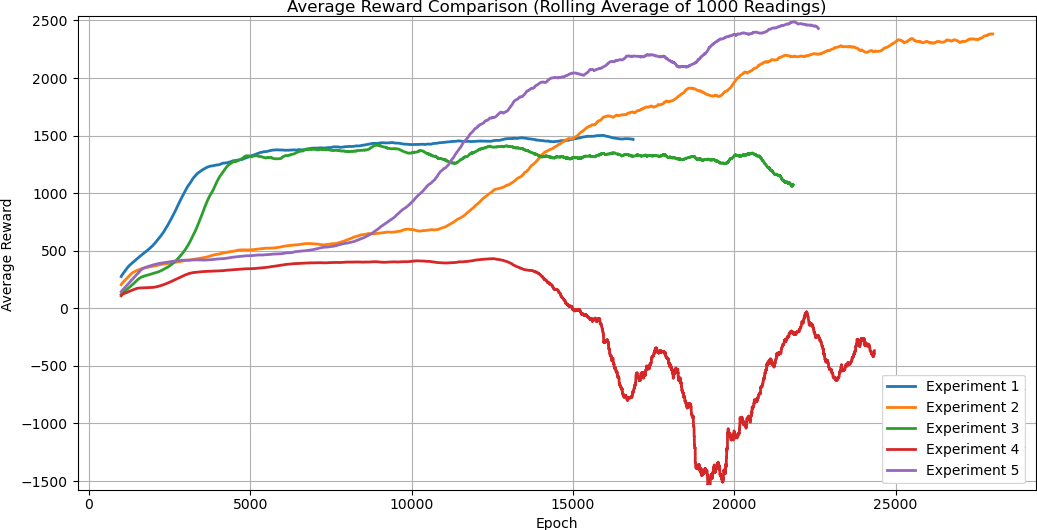}
    \caption{Average Reward Comparison (Rolling Average of 1000 Readings). Refer to \autoref{tab:experiments} for experiment details.}
    \label{fig:Summary}
\end{figure}

\autoref{fig:Summary} provides a comprehensive comparison for different PPO implementations, illustrating the performance impact of different quantization strategies and demonstrating how dynamic standardization and adaptive quantization methods can optimize PPO performance.

\begin{table}[t!]
\caption{Overview of Experiment Attributes}
\centering
\setlength{\tabcolsep}{1pt}
\begin{tabular}{|c|cccc|cc|}
\hline
\multirow{4}{*}{\rotatebox[origin=c]{70}{\shortstack{Experiment\\Index}}} & \multicolumn{4}{c|}{Standardization} & \multicolumn{2}{c|}{\multirow{2}{*}{\shortstack{Uniform\\Quant.}}} \\ \cline{2-5}
 & \multicolumn{3}{c|}{Rewards} & Values & \multicolumn{2}{c|}{} \\ \cline{2-7} 
 & \multicolumn{1}{c|}{\multirow{2}{*}{\shortstack{Dynamic\\Std.}}} & \multicolumn{1}{c|}{\multirow{2}{*}{\shortstack{Block\\Std./De-Std.}}} & \multicolumn{1}{c|}{\multirow{2}{*}{\shortstack{Block\\Std./No De-Std.}}} & \multirow{2}{*}{\shortstack{Block\\Std./De-Std.}} & \multicolumn{1}{c|}{\multirow{2}{*}{\rotatebox[origin=c]{30}{Rewards}}} & \multirow{2}{*}{\rotatebox[origin=c]{30}{Values}} \\
 & \multicolumn{1}{c|}{} & \multicolumn{1}{c|}{} & \multicolumn{1}{c|}{} &  & \multicolumn{1}{c|}{} &  \\ \hline
1 & \multicolumn{1}{c|}{} & \multicolumn{1}{c|}{} & \multicolumn{1}{c|}{} &  & \multicolumn{1}{c|}{} &  \\ \hline
2 & \multicolumn{1}{c|}{\ding{51}} & \multicolumn{1}{c|}{} & \multicolumn{1}{c|}{} &  & \multicolumn{1}{c|}{} &  \\ \hline
3 & \multicolumn{1}{c|}{} & \multicolumn{1}{c|}{\ding{51}} & \multicolumn{1}{c|}{} & \ding{51} & \multicolumn{1}{c|}{\ding{51}} & \ding{51} \\ \hline
4 & \multicolumn{1}{c|}{} & \multicolumn{1}{c|}{} & \multicolumn{1}{c|}{\ding{51}} & \ding{51} & \multicolumn{1}{c|}{\ding{51}} & \ding{51} \\ \hline
5 & \multicolumn{1}{c|}{\ding{51}} & \multicolumn{1}{c|}{} & \multicolumn{1}{c|}{} & \ding{51} & \multicolumn{1}{c|}{\ding{51}} & \ding{51} \\ \hline
\end{tabular}
\label{tab:experiments}
\end{table}

\subsection{Hardware Implementation of \HEPPO}

A parameterized Verilog model of \HEPPO's proposed pipelined architecture, as described in \autoref{sec:Architectural Details of HEPPO}, has been developed with a data width of 32 bits (after fetching and de-quantizing the elements). The AMD-Xilinx Zynq® UltraScale+™ MPSoC ZCU106 Evaluation Kit was chosen to host our implementation. This device integrates a quad-core Arm® Cortex™-A53 processing system (PS) and a dual-core Arm Cortex-R5 real-time processor, providing the necessary computing for running the environment. The FPGA fabric within the Programmable Logic (PL) provides extensive resources for custom logic implementation, including neural network inference and GAE computation.

For the DNN inference within the PL, we adapt the systolic array implementation introduced by Meng et al. (2020) \cite{ppo_paper}. Their design achieves a clock frequency of 285 MHz, whereas our overall system is designed to run at 300 MHz. As all design subsystems operate sequentially (processing does not overlap in time), it's advantageous to enable each subsystem to run at its highest frequency. While this creates multiple clock domains, data synchronization is not required because all subsystems operate sequentially and communicate through BRAMs. However, control signals across domains, such as those indicating that processing has ended and data is ready, still need to be synchronized.

\subsubsection{Area Utilization}

\begin{figure}[b]
\centering
\includegraphics[width=\columnwidth]{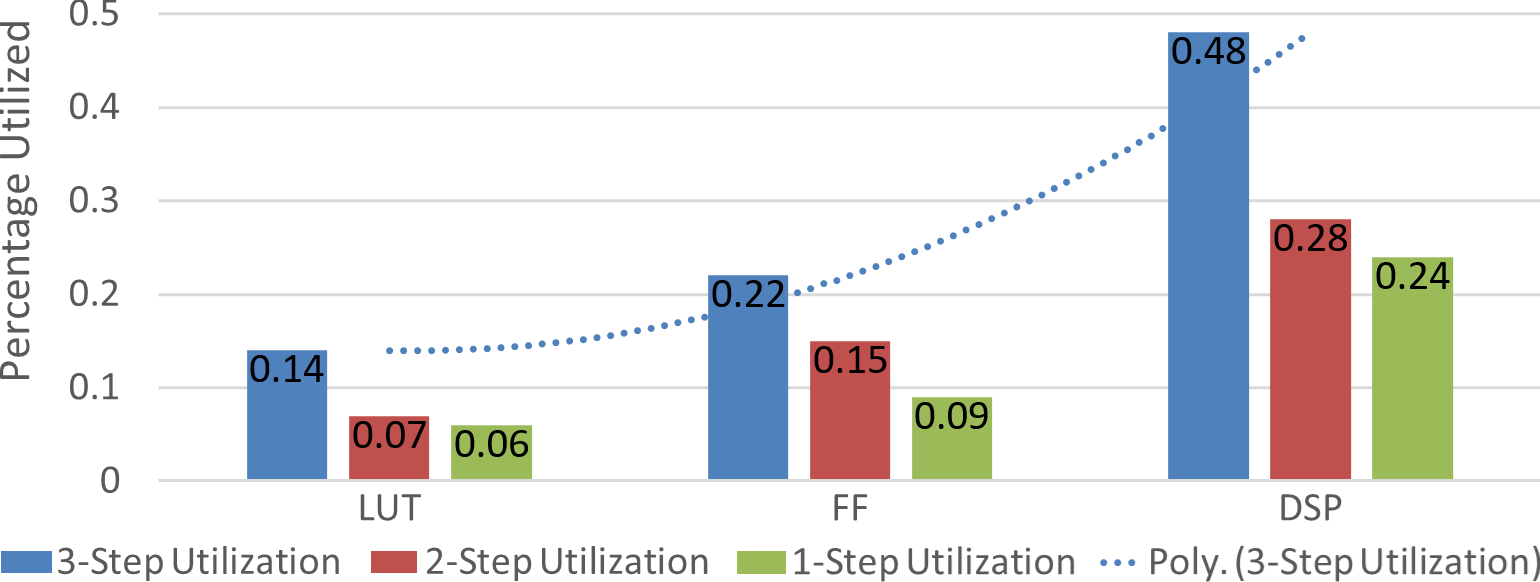}
\caption{Resource Utilization Percentage for n-Step Lookahead PE}
\label{fig:utilization}
\end{figure}

\autoref{fig:utilization} illustrates the resource utilization percentages for 1-step, 2-step, and 3-step lookahead implementations per Processing Element (PE). As seen, there is a quadratic increase in resource usage with each increase in \(n\). This figure highlights how the increase in lookahead steps (n) impacts the utilization of various resources (LUTs, FFs, and DSPs), demonstrating a clear quadratic trend. Based on our implementation, we found that \(n > 1\) allows the system to operate at a maximum frequency of 300 MHz. This is due to the intensive pipelining and absence of pipeline stalls. Hence, a single PE is estimated to handle 300 million elements per second with the continuous data flow supported by the efficient design of the FILO BRAM memory system which ensures the required throughput every cycle. This is in contrast to a normal CPU-GPU system that suffers from DRAM memory access latency, buffering, and scheduling, all of which are a great bottleneck.

\begin{table}[t]
\caption{Resource Utilization for a 2-step lookahead system}
\label{tab:resource_allocation}
\centering
\setlength{\tabcolsep}{1pt}
\begin{tabular}{|l||c|c|c|}
\hline
\textbf{Resource} & \textbf{Total Usage (64 PEs)} & \textbf{Available} & \textbf{Utilization (\%)} \\ \hline\hline
\textbf{LUTs} & 12864 & 274080 & 4.69 \\ \hline
\textbf{FFs} & 54336 & 548160 & 9.91 \\ \hline
\textbf{DSPs} & 768 & 2520 & 30.48 \\ \hline
\end{tabular}
\end{table}

We choose to work with 2-step lookahead. The resource utilization in \autoref{tab:resource_allocation} is estimated for 64 PEs based on our single PE implementation. It shows that the ZCU106 Evaluation Kit can comfortably accommodate our design. The utilization percentages for LUTs, FFs, and DSPs are well within the available resources, with the most significant utilization being DSPs at 17.7\%. This efficient usage ensures that the system can run at the desired frequency and handle the required throughput without encountering resource constraints.

\subsubsection{Memory Utilization Requirements} 

For 64 trajectories and 1024 timesteps, with rewards and values overwritten by advantages and rewards-to-go, the memory required is 128 bytes per timestep, totaling 128 KB for 1024 timesteps.

\textbf{BRAM Utilization:}
Each BRAM provides 36 Kb of storage, so storing 128 KB requires approximately 29 BRAM blocks (around 9\% utilization).

\textbf{Bandwidth Requirement:}
We read 64 rewards and 64 values (128 bytes) and write 64 advantages and 64 rewards-to-go (128 bytes) per clock cycle, requiring a total bandwidth of 256 bytes/cycle.

Each dual-port BRAM handles 4 bytes per port, per cycle. To meet the 256 bytes/cycle requirement, 57 BRAM ports are needed. Since each dual-port BRAM has 2 ports, this translates to 32 BRAM blocks (10\% utilization) required to support both memory storage and bandwidth needs, ensuring efficient parallel processing.

\subsubsection{System Estimated Speedup}

We conducted a test on a standard GAE implementation \cite{yang2020} on a CPU-GPU system comprising 32 cores each is Intel(R) Xeon(R) Silver 4216 CPU @ 2.10GHz, and a Tesla V100-SXM2-32GB GPU, it was concluded that this setup can handle \(\approx 9000\) elements per second. This is interpreted by the nature of this phase which processes trajectories of unequal sizes in reverse, this is traditionally achieved by iterating over one trajectory at a time not in batch form. However, in our implementation, we process a batch of 64 trajectories at a time in custom hardware made specifically for accelerating this phase. Hence, our system can theoretically handle around 2 million times faster than a traditional implementation, significantly reducing the time taken at the GAE stage and increasing the PPO speed by around 30\%.

In addition, having our FILO memory on-chip with the CPU cores as well as the FPGA greatly reduces the memory access time of storing and fetching the trajectories which account for around 11.73\% of the PPO time.

Finally, our proposed solution opens the door for a full PPO system implementation on-device, and it can adapt to any cutting-edge implementation of DNN in the PL. For this research paper, we adapt the systolic array implementation by Meng et al. (2020), leveraging the FPGA’s capabilities to handle high-throughput neural network inference, backpropagation, and GAE computations. Their work is claimed to have achieved substantial performance improvements, ranging from 2.1× to 30.5× when compared to state-of-the-art CPU implementations and 2× to 27.5× when compared to CPU-GPU implementations.

\section{Conclusion and Future Work}

We introduced \HEPPO (Hardware-Efficient Proximal Policy Optimization), an FPGA-based implementation designed to accelerate the GAE stage of the PPO algorithm. \HEPPO utilizes dynamic reward standardization and 8-bit uniform quantization, reducing memory usage by 4x and increasing cumulative rewards by 50\%.

Our innovative memory layout system, using FILO storage and dual-port memory, efficiently handles rewards, values, and advantages, ensuring high throughput and compact data management. The ultra-pipelined Process Element (PE) unit, operating at 300 MHz, greatly enhances throughput and efficiency, outperforming conventional CPU-GPU systems and improving PPO training efficiency by an estimated 30\%.

\HEPPO leverages AMD-Xilinx Zynq UltraScale+ MPSoC’s capabilities, integrating environment simulation, neural network inference, backpropagation, and GAE computation on a single SoC. This minimizes communication latency and optimizes data handling which originally accounted for around 11\% of the training. 

Further incremental optimizations of \HEPPO’s SoC are possible. Overclocking techniques \cite{Brant-FCCM2013} and bit-serial computation \cite{Abdelhadi-FPT2019} in FPGAs can be employed to accelerate overall processing. To mitigate power consumption, multiple clock domains can be implemented for the ARM cores, the DNN, and the GAE calculations. High-performance clock-domain crossing (CDC) FIFOs can facilitate faster data transfers \cite{Abdelhadi-FPL2021,Abdelhadi-FCCM2021,Abdelhadi-NorCas2020,Abdelhadi-NocArc2020,Abdelhadi-ASYNC2017}. Additionally, hardware-efficient data compression methods, optimized for deep learning workloads, can be leveraged to minimize data transfers \cite{Lascorz2024,Zadeh2022,Edo2021}.

Future work should focus on optimizing custom hardware for other phases of the PPO algorithm, particularly in accelerating environment simulation, which consumes 47\% of the training time. Investigating techniques for dynamic High-Level Synthesis of environments on FPGA and implementing loss calculation on FPGA could eliminate the need for CPU cores, significantly boosting the computational efficiency of the algorithm.

\section*{Acknowledgements}
Many thanks to the anonymous reviewers for their valuable feedback and suggestions. This research would not have been possible without access to design tools and libraries provided by the Canadian Microelectronics Corporation (CMC) and access to the Compute Canada Database (CCDB) through the Digital Research Alliance of Canada. This research was funded by the Natural Sciences and Engineering Research Council of Canada (NSERC) Discovery Grant program.


\end{document}